\documentclass[a4paper]{spie}  
\usepackage[]{graphicx}

\def\degr{\hbox{$^\circ$}}

\title{MicroLux: high-precision timing of high-speed photometric observations} 

\author{Felix Hormuth
\skiplinehalf
	Max-Planck-Institut f\"ur Astronomie, K\"onigstuhl 17, 69117 Heidelberg, Germany
}

\authorinfo{Further author information: (Send correspondence to F. Hormuth)\\F. Hormuth: E-mail: hormuth@mpia.de, Telephone: +49 (0)6221 528 221}

\begin{document} 
  \maketitle 

\begin{abstract}
MicroLux is a GPS-based high precision and high speed timing add-on to the Calar Alto Lucky Imaging camera AstraLux. It allows timestamping of individual CCD exposures at frame rates of more than 1 kHz with an accuracy better than one microsecond with respect to the UTC timeframe. 
The system was successfully used for high speed observations of the optical pulse profile of the Crab pulsar in January and November 2007. 
I present the technical design concept of MicroLux as well as first results from these observations, in particular the reconstructed pulse profile of the pulsar\footnote{Based on observations collected at the Centro Astron\'omico Hispano Alem\'an (CAHA) at Calar Alto, operated jointly by the Max-Planck Institut f\"ur Astronomie and the Instituto de Astrof\'isica de Andaluc\'ia (CSIC).}.
\end{abstract}



\section{INTRODUCTION}

High spatial or spectral resolution observing techniques are nowadays a common standard 
at large modern optical telescopes. 
However, only few instruments exist that explore the temporal dimension with resolution
higher than a few 10\,Hz. 
On one hand, classical CCD imaging cameras have typical readout times of several ten seconds, resulting in increasingly small duty cycles if high temporal resolution is demanded.
Cameras with fast readout capabilities on the other hand, e.g. frame transfer CCDs, suffer from
increased readout noise, limiting the minimum brightness or exposure time significantly.

Nevertheless, high temporal resolution at optical wavelengths has its applications, especially
in the near infrared (NIR), where sky brightness dominates the noise budget even at short
exposure times. High speed photometry of stellar occultations by the moon, planets, or
planetary rings and satellites, is a standard observing technique since several decades now.
Typical instruments built for this purpose at visible wavelengths are based on photon counting 
Avalanche photodiodes\cite{Kanbach:2003,Stefanescu:2008} or conventional CCDs using
high readout clocks\cite{Dhillon:2007}.

With the advent of electron multiplying CCDs (EMCCD) with virtually zero readout noise and single-photon
detection capabilities at wavelengths below 1\,$\mu$m, high speed photometry of faint targets in 
the visible is now technically feasible even with medium sized telescopes, and with minimum
technical effort.
The Lucky Imaging camera AstraLux\cite{Hormuth:2008} at the Calar Alto 2.2-m telescope is such 
a system. Originally designed for high spatial resolution observations, the frame transfer EMCCD
can be read out with frame rates of up 1.5\,kHz, depending on binning and windowing settings.
If all 512$\times$512\,pixels of the 24$\times$24\,arcsec large field of view are read unbinned,
the resulting frame rate is still $\approx$34\,Hz.

Unfortunately, high temporal resolution alone without precise knowledge of the absolute times of the individual measurements is not always sufficient. 
While the fringe pattern observed during a stellar occultation can be used to determine the star's 
diameter on the basis of the known frame rate only, absolute timing is needed for precision 
measurements of e.g. eclipsing binary star parameters or the radio-optical delay of a pulsar.
Decades ago, the signals of terrestrial navigation systems or television stations were used
as time and frequency standards\cite{Papaliolios:1970,Lohsen:1981}. Today, the 
Global Positioning System (GPS) satellite network is a powerful and reliable tool for absolute time
measurements, and in fact GPS based timing solutions are employed in modern high
speed photometry instruments.

\begin{figure}[htb!]
	\begin{center}
   		\begin{tabular}{c}
   			\includegraphics[width=20.5cm,angle=90]{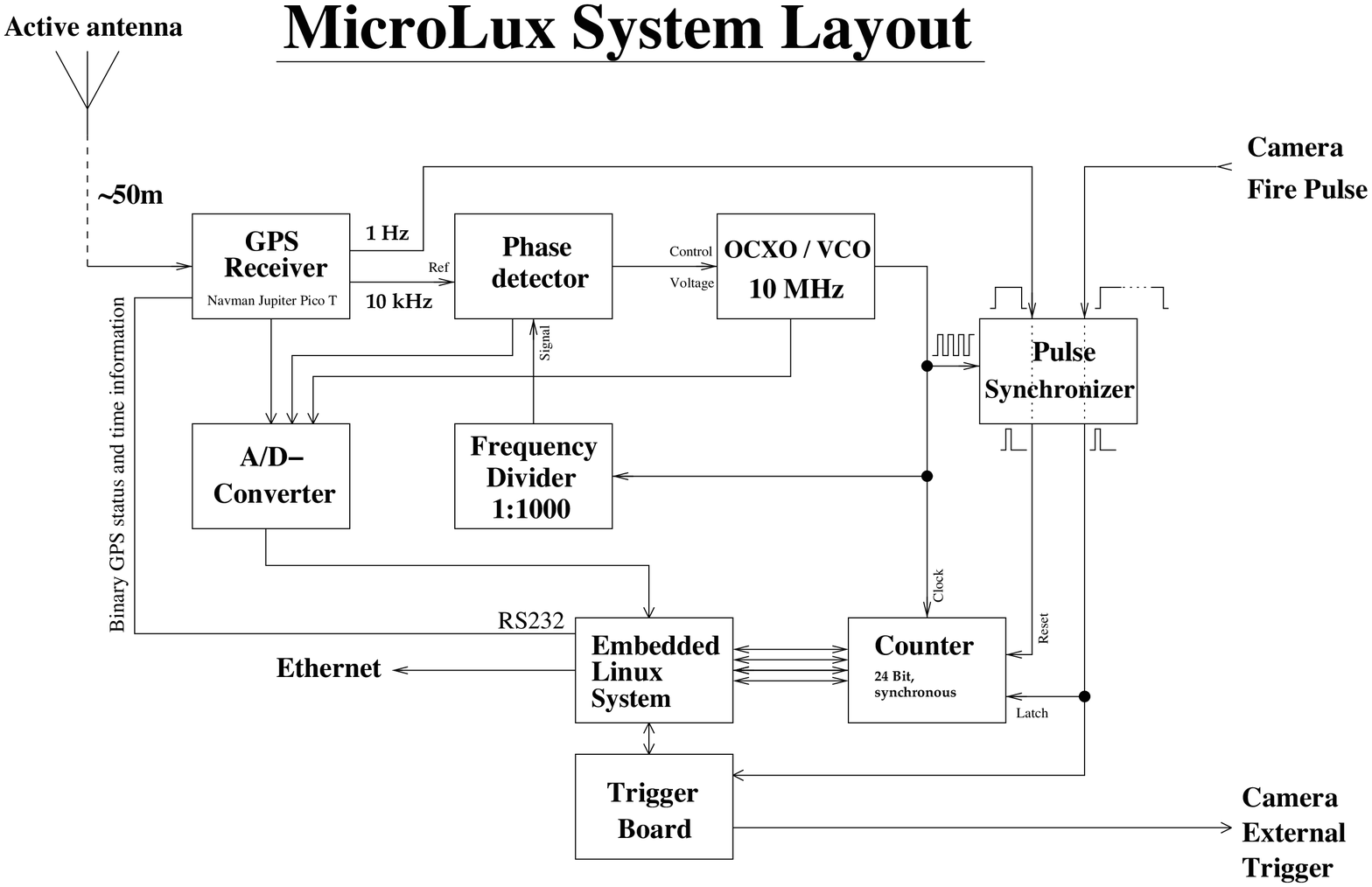}
 	 	 \end{tabular}
 	  \end{center}
  	\caption[example] 
		{ \label{fig:layout} 
	Block layout of the MicroLux timing system. See text for further description of the individual
	components.
	}
\end{figure} 

\begin{figure}[htb!]
	\begin{center}
   		\begin{tabular}{c}
   			\includegraphics[width=12cm,angle=0]{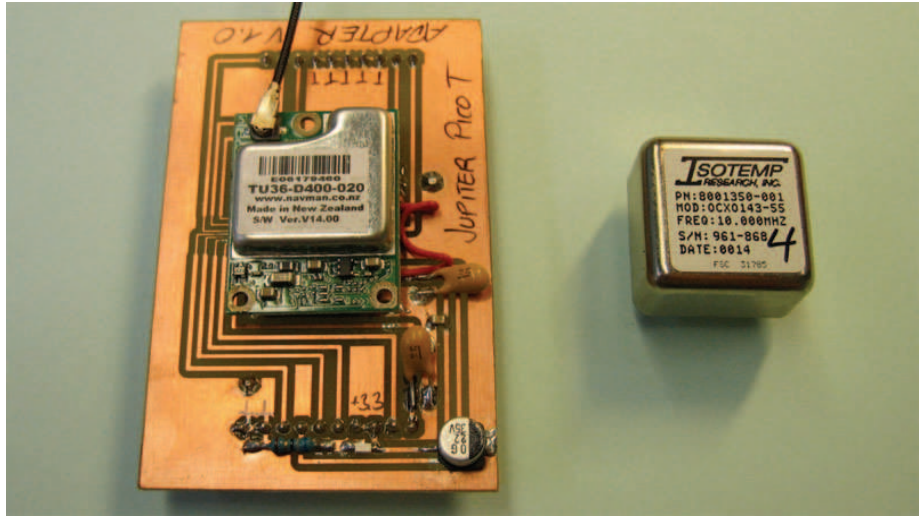}
 	 	 \end{tabular}
 	  \end{center}
  	\caption[example] 
		{ \label{fig:gpsocxo} 
	 The Jupiter Pico T GPS receiver on its adapter board and the Isotemp OCXO 143-55
	 module used in MicroLux.	}
\end{figure}

\begin{figure}
	\begin{center}
   		\begin{tabular}{c}
   			\includegraphics[width=5.7cm,angle=270]{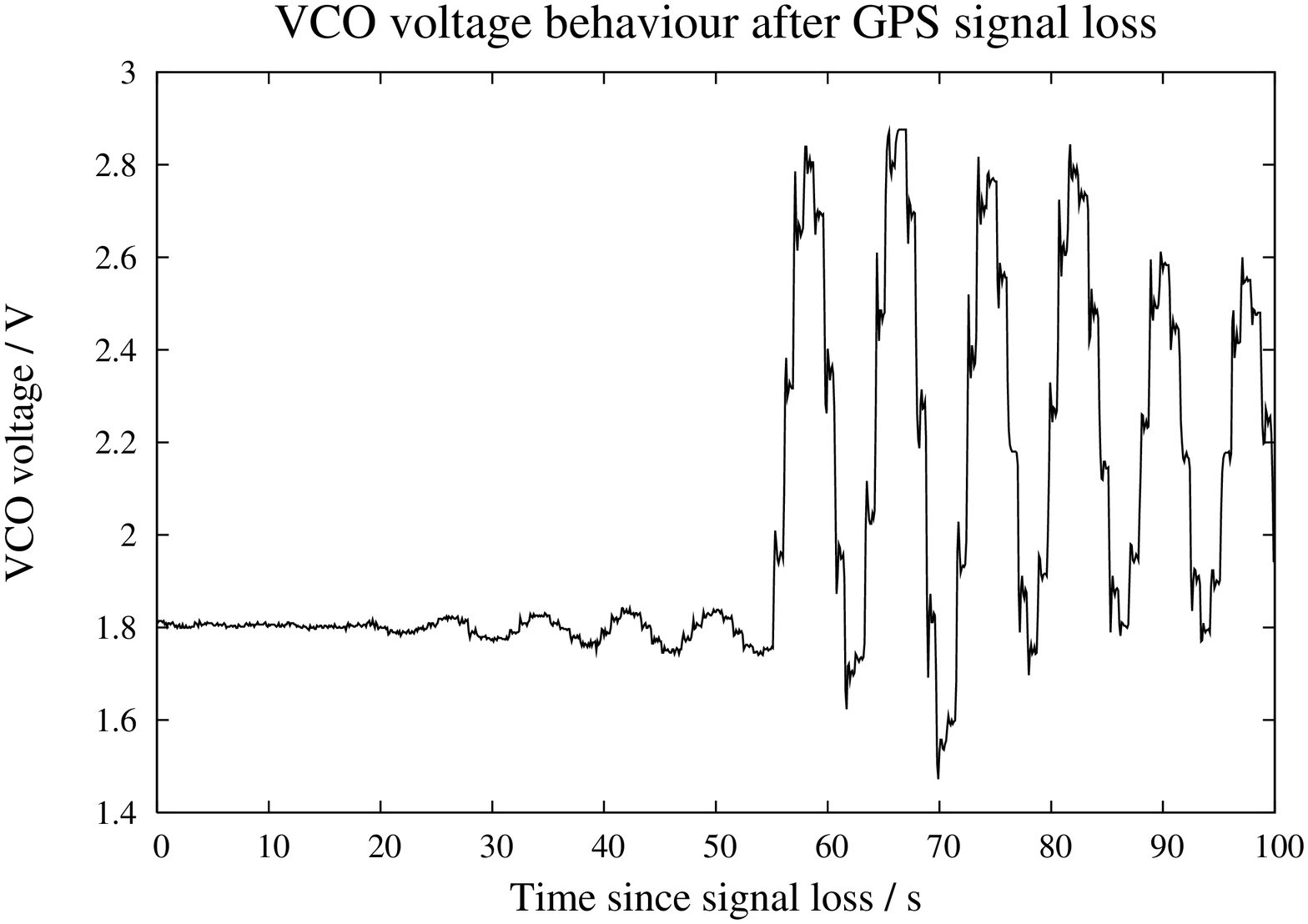}
   			\includegraphics[width=5.7cm,angle=270]{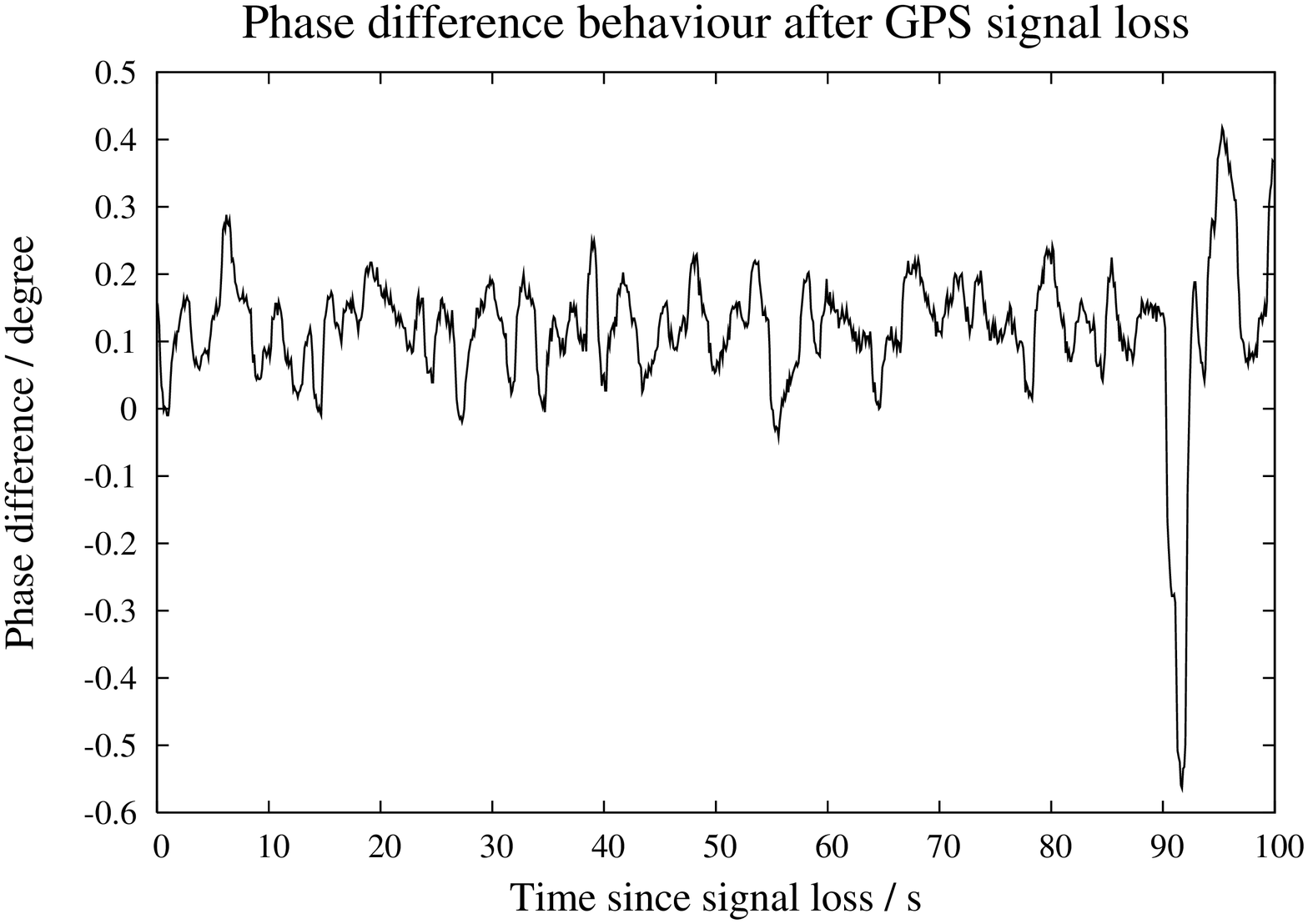}			
 	 	 \end{tabular}
 	  \end{center}
  	\caption[example] 
		{ \label{fig:vco} 
		Measurements of the VCO voltage and the phase difference between the PLL signals
		after disconnection of the GPS receiver's antenna at 20\,s after measurement start.
		While the PLL maintains perfect synchronisation for more than a minute, the loss of
		satellite signal is clearly indicated by oscillations of the VCO voltage after only
		few seconds. Monitoring the VCO voltage and its first derivative is thus an extremely
		sensitive tool for the detection of GPS signal loss conditions.
	 	}
\end{figure} 

Commercially available devices for GPS time tagging often allow for only low frame
rates, not suitable for very high speed photometry, or were specifically designed for
complex synchronisation tasks, and are quite expensive.
In order to enable photometry at more than 1\,kHz frame rate with sub-microsecond accuracy
and at an affordable price, the MicroLux system was designed to enable precision time-tagging
of individual frames taken with the AstraLux Lucky Imaging camera.

\section{INSTRUMENT LAYOUT}
The layout of the AstraLux hardware is shown in Figure\,\ref{fig:layout}. The heart of the system is
a commercial GPS receiver board, model Jupiter~Pico~T from Navman Ltd., UK (see Figure~\ref{fig:gpsocxo}). After initial reception
of at least four GPS satellites, the signal of a single satellite is sufficient to maintain a stable 1\,Hz
and 10\,kHz signal, aligned to each other and with respect to the UTC second with an accuracy 
better than 30\,ns. The Pico\,T module was especially designed for timing applications and
assumes to be at a fixed location after establishing the initial 3D position.

Since MicroLux is physically located at the mirror cell of the telescope, but proper GPS reception
is only possible outside the dome, a standard active GPS antenna is connected via 50\,m of RG\,213
coaxial cable and placed at one corner of the telescope building's catwalk. This is sufficient to deliver
good signal levels of usually 6--8 satellites not blocked by the metallic dome structure.

The GPS module's 10\,kHz output is further used to discipline the frequency and phase of an
ovenised crystal oscillator (OCXO), model OCXO 143-55 by Isotemp Corp. (see Figure~\ref{fig:gpsocxo}), via a phase
locked loop (PLL) circuit.
The OCXO alone is with a typical short-time frequency error of 1\,Hz already very stable,
and the PLL can thus work with a long time-constant of 10\,s. Together with the time
needed for warm-up of the OCXO and for the position fix of the GPS, the typical time until
PLL lock is $\approx$4\,min after system power-on. Given that MicroLux is switched on
usually only once per observing run and then left running, this is more than acceptable.

For microsecond accuracy it would in principle be possible to use the OCXO tuned to
10\,MHz in un-disciplined mode, since the frequency error would only sum up to a
timing error less than 1\,$\mu$s until the next 1\,Hz pulse of the GPS. However, by controlling the OCXO
via the PLL, the VCO voltage becomes a powerful diagnostic tool. Any deviations
from the regular mean value or increased noise are strong indicators for problems
with the GPS's timing accuracy, e.g. caused by a total loss of the satellite signal 
(see Figure~\ref{fig:vco}) .
 
The GPS disciplined 10\,MHz signal feeds a synchronous 24-Bit counter module, which is
reset at every full UTC second by the GPS's 1\,Hz signal. At any given moment the counter
value thus corresponds to the fraction of UTC second in units of 100\,ns.
An external signal, in this case the ``Fire'' pulse of the camera, rising at exposure start,
latches the current counter value into three 8-Bit registers. The system's control computer
is notified via one of its I/O lines, reads out and saves the time tag of the last trigger event.

The control computer is an embedded Linux system running Kernel version 2.6, model ``Foxboard LX'' from Acme Systems, Italy. It provides various I/O lines, up to four serial ports, two USB ports, and a 100\,MBit Ethernet interface. It establishes the connection between the counter hardware and the
observatory's computer network. A simple command line software is used to start time-tagging of a
predefined number of events. 

An 8-channel A/D converter with 12 Bit resolution is interfaced to the Linux board to allow monitoring
of supply voltage levels, PLL parameters (VCO voltage, phase difference, oven current), and
temperature in the system's enclosure. Together with the GPS module's status information 
received via one of the RS232 ports this provides full information about system health, GPS reception
quality, and PLL lock status.

The CCD exposure start signal (the ``Fire'' pulse) is not directly fed into the counter module or
Linux system, but first aligned with the 10\,MHz signal and shortened to a length of 50\,ns
to avoid latch-up conditions and runt pulses in the subsequent stages. After passing this
``Pulse Synchroniser'' module it is fed into the trigger board. This is basically a flip-flop circuit,
used to latch the exposure start signal and to indicate overflow conditions if a new start signal
is received before the control computer was able to read out the timing information of the
last one. This board also provides an output which can be used to trigger individual exposures
of the camera or to start a free-running exposure sequence at a predefined time.

Special emphasis was put on using high-quality, linear low-noise power supplies for the GPS and
PLL part, decoupled from the power supplies of the counter and Linux board. Together with the
long time constant of the PLL, this leads to low phase noise of the GPS's 10\,kHz and the
disciplined OCXO's 10\,MHz signals.

\begin{figure}
	\begin{center}
   		\begin{tabular}{c}
   			\includegraphics[width=16cm,angle=0]{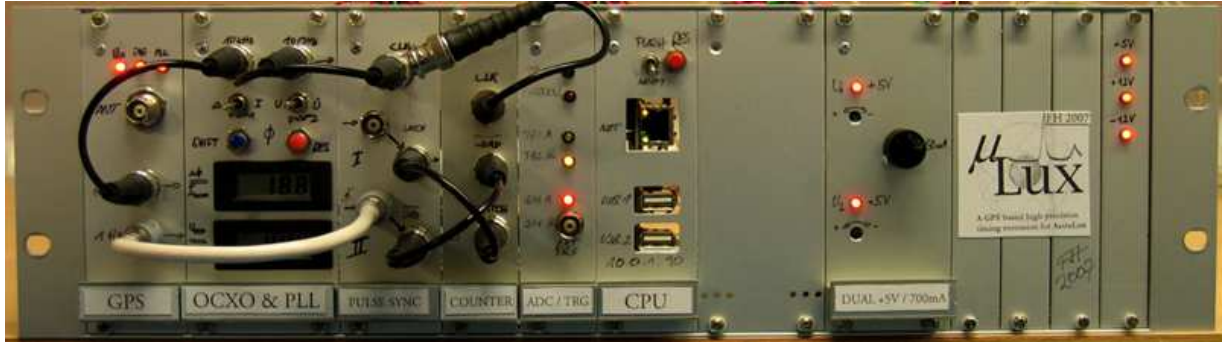}
 	 	 \end{tabular}
 	  \end{center}
  	\caption[example] 
		{ \label{fig:muluxclose} 
	 	The MicroLux 19-inch electronics rack in full configuration. From left to right: GPS module, 
		disciplined 10\,MHz OCXO and PLL, pulse shaper, synchronous counter module, 
		trigger board, embedded Linux system, and power supply modules.}
\end{figure}

The complete system, shown in Figure\,\ref{fig:muluxclose}, fits into a standard 19\,inch electronics
subrack, mounted in the electronics rack of the AstraLux camera. The weight is approximately
5\,kg. Total hardware costs were below 1000\,Euro, and the system could be developed and built 
by a single person in two months.

\subsection{Software}
A cross-platform development kit allows  to write software for the Foxboard control computer on
any Linux PC. The compiled binary files are copied via FTP or SFTP to the 
target system. 

The MicroLux software consists of a simple command line utilitiy.
The program is invoked with the number of frames that will be acquired by the camera
and a name for the logfile to be generated. Before any frames are acquired, the MicroLux
software will synchronise itself with the UTC second pulses of the GPS system and obtain
the full time and date information by decoding the binary data received from the GPS 
serial port. After synchronisation, the software waits for the next full UTC second to issue
a start signal via the trigger board. This signal starts the free-running time-series
acquisition of the AstraLux camera. 

Each time the trigger board indicates the start of a new frame, the software reads out
the counter value, i.e. determines the precise start time of the frame, and stores
this information to the control computer's RAM. It is checked and recorded if the trigger 
board has indicated an overrun condition. 

Upon completion of the time series, all frame start times are written to the RAM file
system, ready for retrieval via FTP or SFTP. A system health logfile is created, 
containing system voltage levels and GPS status information for a timespan of
10\,s. Both files can be processed with own IDL programs, allowing to assess 
the quality of the timing information and to check for any anomalies.

A second command line tool continuously decodes the binary GPS status information
and reads the A/D converter voltages. The values are displayed on the terminal screen,
and any deviations from pre-defined expectations (e.g. due to problems with one of the
power supply voltages, or anomalous GPS position measurements) are highlighted.
This tool is mainly used during setup or at the begin of observing nights to get a quick
and informative overview of the system's health condition.

At the moment, this diagnostic tool cannot be run in parallel to the time-tagging software,
since this has to be executed with high scheduler priority to allow time-tagging at kilohertz
frame rates. A future version of MicroLux will use an FPGA circuit replacing the discrete
counter and register module, employing a first-in-first-out storage register for up
to 100 time-tagging measurements, leading to less strict demands on the Linux board's
real-time capabilities.

\begin{figure}
	\begin{center}
   		\begin{tabular}{c}
   			\includegraphics[width=5.7cm,angle=270]{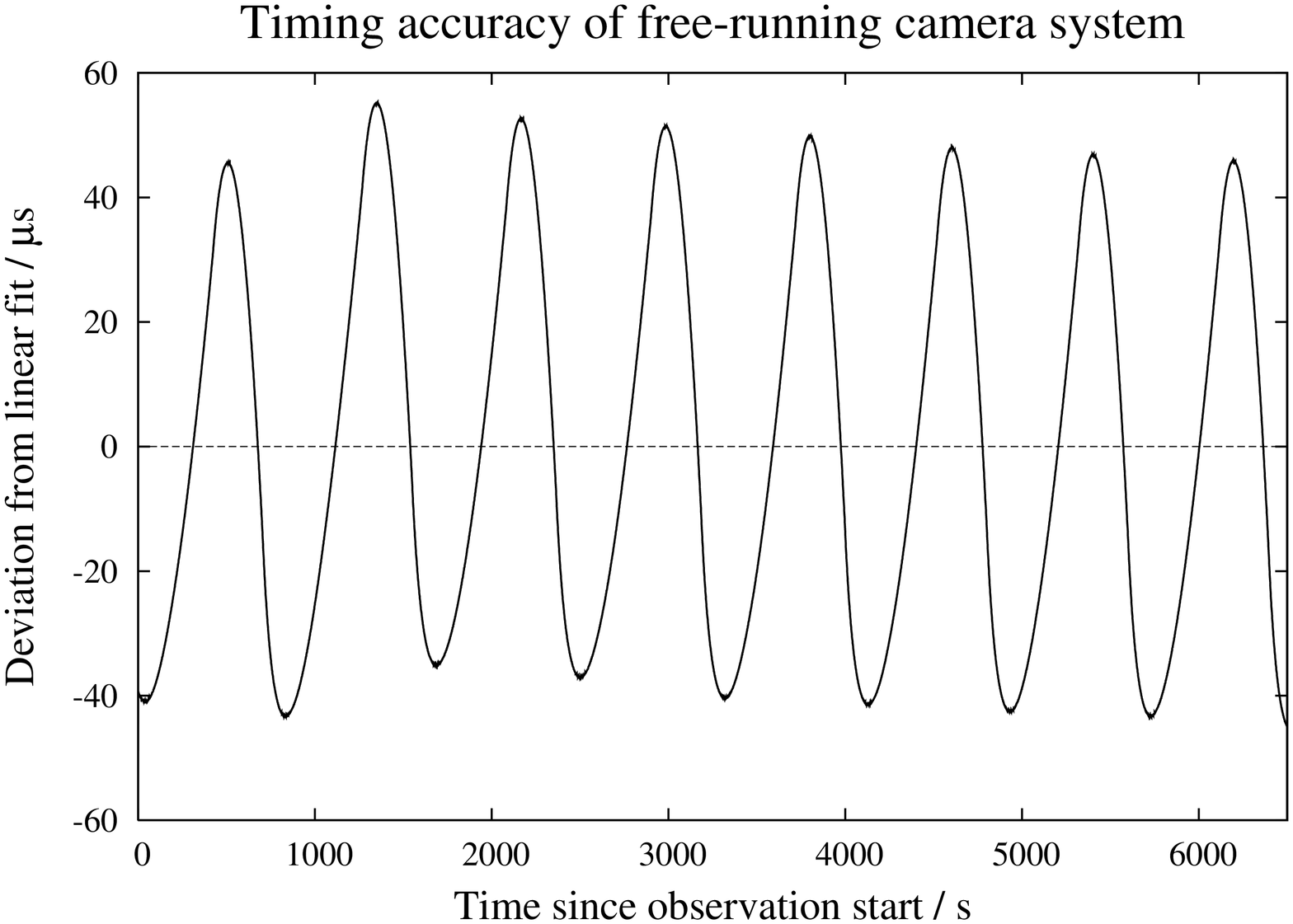}
   			\includegraphics[width=5.7cm,angle=270]{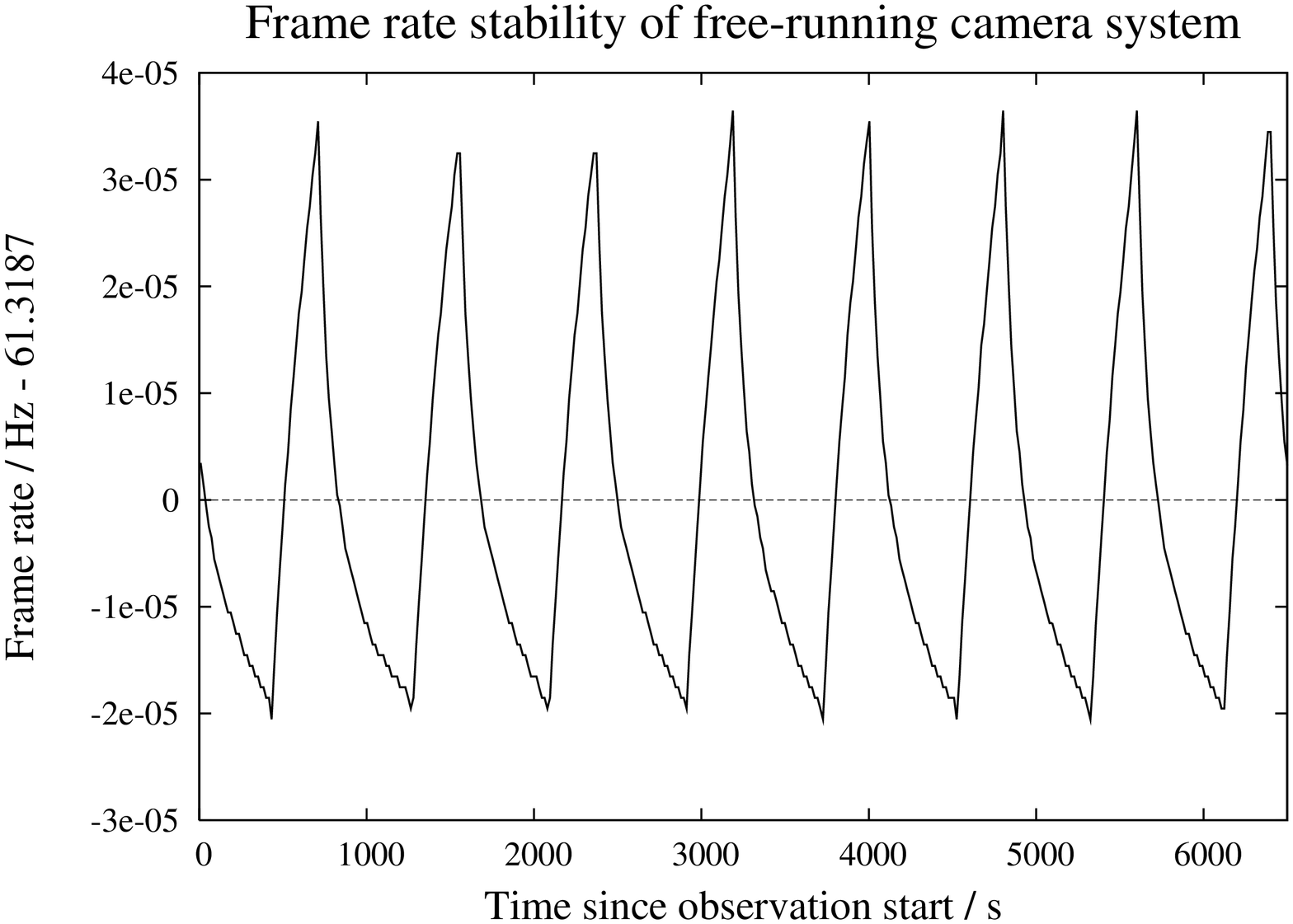}			
 	 	 \end{tabular}
 	  \end{center}
  	\caption[example] 
		{ \label{fig:timejitter} 
			Frame rate stability of the free-running AstraLux camera system
			
			{\em Left:} Deviation of the actual frame start from a linear fit to the
					measured start times. The frame rate was $\approx$61\,Hz.
					The deviations of up to 50\,$\mu$s are a feature of the
					camera, and not introduced in the timing system.
			
			{\em Right:} Measured frame rate for the same data set.
	 	}
\end{figure} 

\begin{figure}
	\begin{center}
   		\begin{tabular}{c}
   			\includegraphics[width=8.5cm,angle=0,trim=1cm 0cm 0cm 1cm]{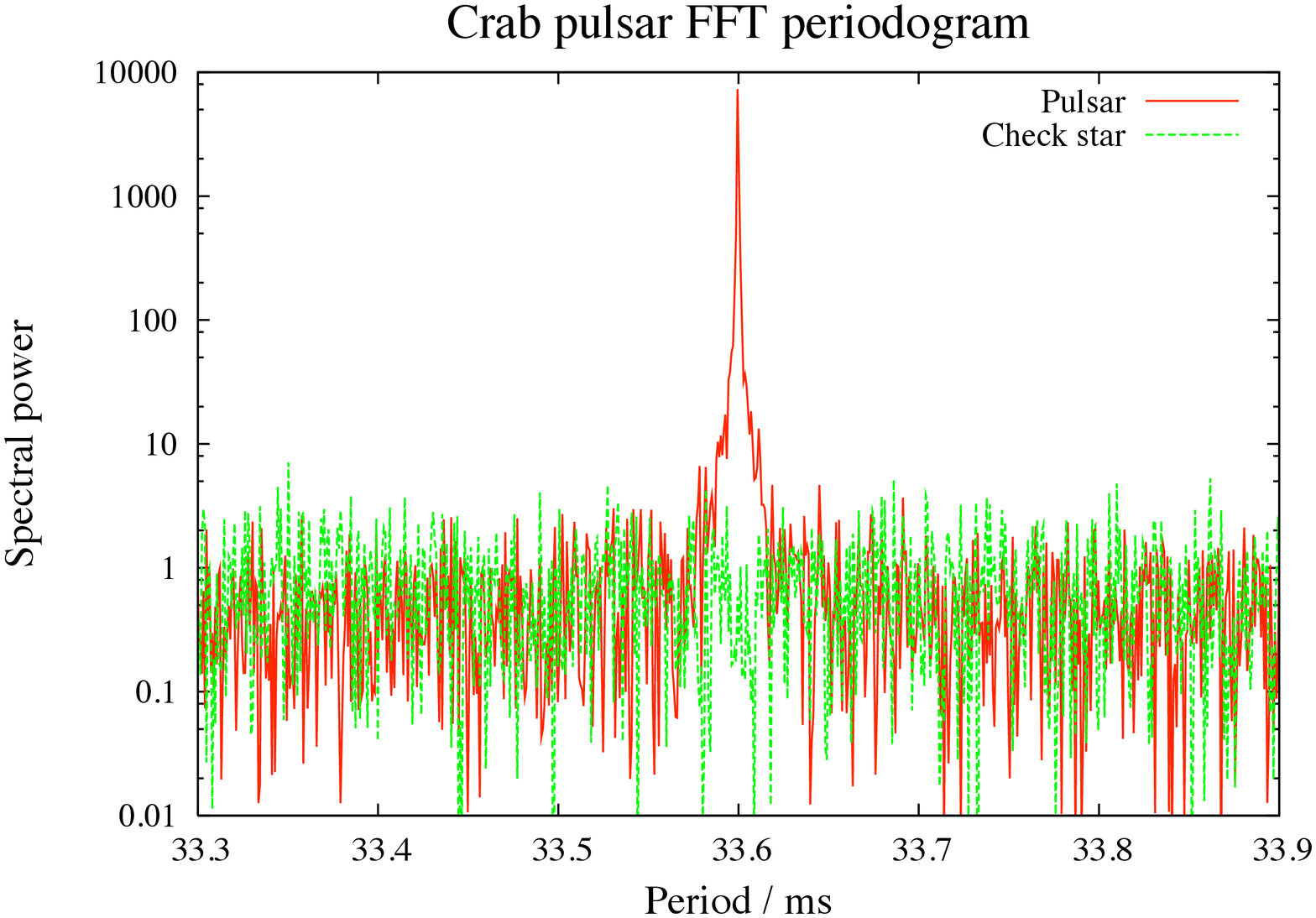}		
   			\includegraphics[width=8cm,angle=0]{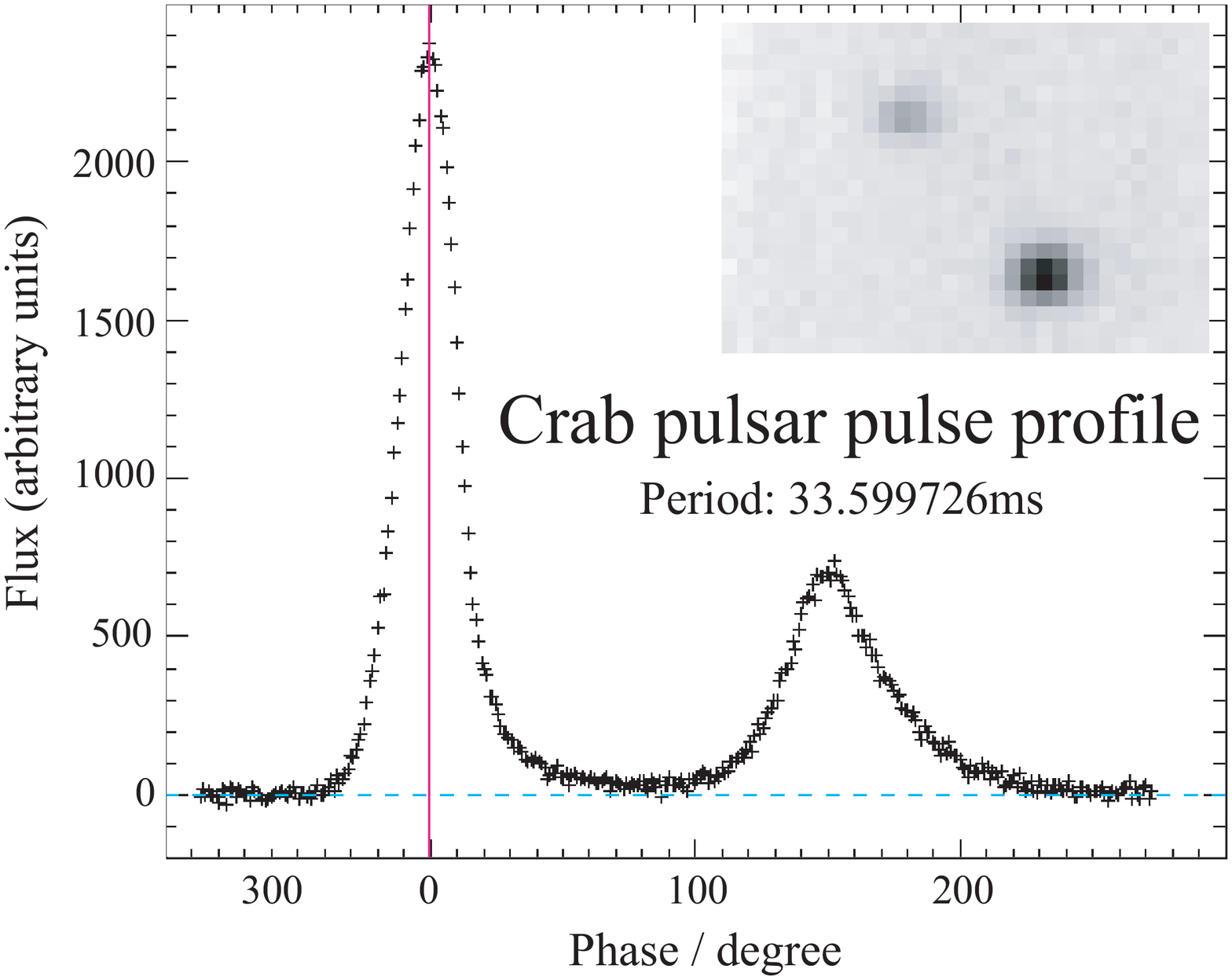}
 	 	 \end{tabular}
 	  \end{center}
  	\caption[example] 
		{ \label{fig:crabpulse} 
		{\em Left:} FFT power spectrum of the raw data, in red for the Crab pulsar and in
		green for the comparison star. Though the pulsar is not visible in single exposures,
		the periodic signal is easily detected with $\approx$5000\,$\sigma$.
		
	 	{\em Right:} The {\em V}-band pulse profile of the Crab pulsar, as seen by AstraLux \& MicroLux.
		The reconstructed profile is based on 700000 images with a single frame exposure 
		time of 1.6\,ms. The inset shows the pulsar at maximum brightness. In long-exposure
		images, the star to the north-east would appear considerably brighter than the pulsar.	}
\end{figure} 

\section{MEASUREMENTS AND OBSERVATIONS}
\subsection{Performance Checks}
Before MicroLux was deployed at the telescope, laboratory tests were conducted
to assess the reliability and accuracy of the system. A second GPS receiver, based on
a Navman Ltd. Jupiter~12 module was used to generate trigger signals with 
precisely known frequencies. With a typical timing RMS of less than 50\,ns, this module
is comparable to the one used in MicroLux, and was actually used in an earlier version
of the system.
Up to 700000 time measurements of trigger signals with frequencies between 10\,Hz and 1000\,Hz,
generated with the Jupiter~12 module, were acquired with MicroLux. This is the maximum
number of events for which the timing information fits into the memory of the Foxboard Linux computer.
In no case the difference between any two trigger signals in these series differed by more
than 100\,ns from the predicted value. This is the period length of the counter clock, and hence
the expected measurement jitter. The statistical component of the measurement error
can hence be assumed to be less than 0.1$\mu$s. Including all delays in the electronics
between signal input and counter module, the systematic delays were estimated as $\approx$300\,ns.
Using the 1\,Hz second pulse as trigger signal, this delay was determined to be in fact 220$\pm$20\,ns.
Even without a systematic correction of this delay by the software, the achieved timing acuracy of
the MicroLux system is hence better than 1\,$\mu$s, the value adopted in the initial 
design considerations.

Tests with the AstraLux camera as source of the trigger pulses revealed that the frame
rate as predicted by the camera software can differ by up to 0.5\% from its actual value.
The frame rate shows high stability, though, with quasi-periodic variations of the 
order of 10\,ppm on scales of minutes (see Figure~\ref{fig:timejitter}). 
They are probably related to temperature 
changes in the camera control computer housing, causing drifts of the camera control
board's master clock frequency. The periodicity most likely reflects the behaviour of the
thermostat regulating the speed of the computer's fan.

This shows that even for applications where absolute timing is not necessary, the MicroLux
system is a valuable tool to determine at least the real frame rate. In the case of precise pulsar
timing applications, measurements of the frame rate variations and the absolute frame
start times are crucial.

\subsection{Observations of the Crab Pulsar}
The Crab pulsar, PSR~J0534+2200, has a period of $\approx$34\,ms at an average
optical magnitude of $V$=16\,mag. Time-resolved observations of its pulse profile were
seen as an optimal test case for AstraLux\,\&\,MicroLux, making full use of the camera's
single photon detection capability and the timing system's accuracy. 

In January 2007, several millions of short exposure frames with integration times 
between 50\,$\mu$s and 2\,ms were acquired in the $B$, $V$, $R$, and $I$ filter, as well 
as unfiltered.  Though this data is not completely reduced yet, early on-site analyses
have proven that the instrument is able to reconstruct the pulse profile and to reliably reproduce
the pulsar period as predicted by radio observations. The data will ultimately allow 
measurements of the radio-optical delay and its dependency on wavelength.
Figure~\ref{fig:crabpulse} shows a Fourier power spectrum of the photometry and 
the reconstructed pulse profile, generated from 700000
$V$-band images with 1.6\,ms single frame exposure time. 

Before the pulse profile could be reconstructed, the pulsar period including
corrections due to pulsar spindown and Doppler effects from earth rotation and 
earth's orbital motion had to be determined to group the single images into phase bins. 
Instead of using a theoretical prediction,
the period was measured in the optical data itself by using a phase dispersion
minimisation (PDM) technique after a first guess of the period by FFT analysis. 

The adopted period length was 0.033599726$\pm$10$^{-9}$\,s. The radio-based
theoretical prediction for the time and location of this observation was kindly
computed by Michael Kramer of Jodrell Bank's pulsar group: 0.03359972563\,s.
The difference between the two values is only 0.4\,ns, well within the error bar
estimated from the PDM algorithm. This corresponds to a relative accuracy of
10$^{-7}$, or a phase error of $\approx$1\degr\ for the full 33600 pulsar periods
that were covered by the data set used in this example. 
This result impressively confirms that MicroLux performs well and provides a valuable extension of AstraLux's capabilities.

\section{CONCLUSIONS}
At total hardware costs of less than 1000\,Euro and within less than two months of development
time it was possible to design and built MicroLux, a high-speed, high-precision time-tagging
system. Using a GPS disciplined ovenised crystal oscillator, MicroLux can determine
the start times of individual CCD exposures with an accuracy better than 1\,$\mu$s at
frame rates of more than 1000\,Hz. Originally built for the Lucky Imaging camera AstraLux at
the Calar Alto 2.2-m telescope, MicroLux could be used with other image acquisition systems
as well thanks to its universal and modularised design.

MicroLux has been used mainly for observations of the optical pulse profile of the
Crab pulsar up to now, demonstrating that the intended design goals could be fully
reached. Further analysis will make use of the absolute timing capabilities of MicroLux
with respect to the UTC timeframe to determine the pulsar's radio-optical delay and its
dependency on wavelength.\\
Other applications include precise timing of binary star eclipses or stellar occultations
by the moon, planets, planetary satellites, or asteroids.

Further planned improvements of MicroLux aim at increased versatility, higher maximum
supported frame rates, and miniaturisation of the system, hopefully leading to a final
design suitable for field-operation, e.g. in the case of stellar occultations by asteroids,
where multiple observers with timing capabilities are needed to gather full information.
The relatively simple design, based mainly on commercially available parts, and the 
low price are certainly great benefits in this respect.



 

\bibliography{microlux}   
\bibliographystyle{spiebib}   

\end{document}